## Research Article

# Shear Thickening in Concentrated Soft Sphere Colloidal Suspensions: A Shear Induced Phase Transition

**Joachim Kaldasch,[1] Bernhard Senge,[1] and Jozua Laven[2]**

[1]*Technische Universität Berlin, Königin-Luise-Strasse 22, 14195 Berlin, Germany*
[2]*Eindhoven University of Technology, P.O. Box 513, 5600 MB Eindhoven, Netherlands*

Correspondence should be addressed to Joachim Kaldasch; joachim.kaldasch@international-business-school.de

A model of shear thickening in dense suspensions of Brownian soft sphere colloidal particles is established. It suggests that shear thickening in soft sphere suspensions can be interpreted as a shear induced phase transition. Based on a Landau model of the coagulation transition of stabilized colloidal particles, taking the coupling between order parameter fluctuations and the local strain-field into account, the model suggests the occurrence of clusters of coagulated particles (subcritical bubbles) by applying a continuous shear perturbation. The critical shear stress of shear thickening in soft sphere suspensions is derived while reversible shear thickening and irreversible shear thickening have the same origin. The comparison of the theory with an experimental investigation of electrically stabilized colloidal suspensions confirms the presented approach.

## 1. Introduction

Concentrated colloidal dispersions are of technological relevance for various industrial products such as paints, pharmaceuticals, cosmetics, lubricants, and food. They are often processed at high shear rates and stresses. In strong viscous flows, colloidal dispersions exhibit a unique transition related to an increase of the shear viscosity, termed shear thickening [1]. This effect makes it harder to pump suspensions and can cause equipment damage and failure in the production processes (flow blockage). Next to reversible shear thickening, also the irreversible aggregation of the dispersion after shear thickening has been reported known as irreversible shear thickening [2]. A fundamental understanding of the relation between particle and flow properties of dense colloidal suspensions is required in order to provide clues minimizing undesired effects or exploiting useful applications of shear thickening [3].

It has long been an issue to understand the underlying mechanisms involved in shear thickening. A number of theoretical attempts were made explaining shear thickening in sheared suspensions of Brownian particles where inertial effects can be neglected (non-Brownian particles [4] and deformable spheres like emulsions, etc. are not considered here). We have to distinguish between hard sphere suspensions with an interaction potential confined to the bare particle (Born-) repulsion and soft sphere suspensions with a combination of van der Waals attraction and long-range electrostatic or steric repulsion in addition to the Born-repulsion.

The dynamics of sheared hard sphere suspensions can be studied in simulations. In dense suspensions, the particle interaction is essentially determined by lubrication forces. When particles are pushed together along the compression axis of a sheared suspension, they must overcome the viscous drag forces between neighbouring particles in order to move away from each other. Above a critical shear rate, particles stick together generating the formation of shear induced clusters (hydroclusters) [5–8]. At high volume fractions, these transient touched particles induce jamming and an increase of the viscosity at a critical shear stress [9]. An alternative concept explaining shear thickening is related to an order-disorder transition. In this approach, an ordered, layered structure of colloidal particles in a sheared suspension becomes unstable above a critical shear rate [10–13]. Shear thickening occurs when lubrication forces between neighbouring particles cause particles to rotate out of alignment of the sheared structure, destabilizing the flow. A more sophisticated approach is based on an ad hoc mode-coupling model. It describes the instability as a stress-induced transition into a jammed state [14–17]. Also suggested is that dilation of confined suspensions may cause a rapid increase of the viscosity termed discontinuous shear thickening [18].

The presented theory is confined to the study of Brownian soft sphere suspensions. As long as the attractive interaction can be neglected, soft spheres can be approximated as so-called effective hard spheres and the abovementioned mechanisms for shear thickening in hard sphere suspensions may apply. However, soft sphere particles have a much more complex interaction than hard spheres. With increasing volume fraction van der Waals attraction causes a coagulation of colloidal particles, in particular when the repulsive stabilization of the particles is small. For soft sphere suspensions, the hard sphere approach to shear thickening has to be extended. An approach aiming at understanding shear thickening in soft sphere suspensions is an activation model [19–21]. It takes advantage from the complex interaction potential and suggests that colloidal particles arranged along the compression axis of a sheared suspension may overcome the mutual repulsion at a critical shear stress. As a result, the viscosity increases when clusters of coagulated particles are formed. For a sufficiently strong attraction between particles, these clusters cannot be disrupted by the applied shear leading to irreversible shear thickening. Otherwise, shear thickening is reversible.

While the activation model is a microscopic approach, this paper presents an alternative mesoscopic approach to shear thickening in dense soft sphere suspensions. It suggests that shear thickening is related to a shear induced phase transition. The transition is formulated in terms of a thermodynamic standard concept known as the Landau model [22, 23]. The Landau model is a mean field theory originally developed to understand symmetry breaking phase transitions. It is a widely accepted theory utilized in particular to model the dynamics of fluids and polymers [24]. Here the theory is used to describe the equilibrium coagulation transition of soft spheres. Taking the coupling of density fluctuations to the viscoelastic medium into account, it can be shown that sheared dense suspensions induce clusters of coagulated particles in equivalence to the microscopic activation model.

The paper is organized as follows. First, shear thickening in suspensions of hard and soft spheres is discussed. After establishing a hydrodynamic model for a dense suspension and deriving a Landau model for the coagulation transition, the models are combined in a subcritical bubble theory to predict the occurrence of coagulated particle clusters. The application of the subcritical bubble approach to sheared dense suspensions allows an estimation of the critical stress for shear thickening. After comparing the model with an experimental investigation, the paper ends with concluding remarks on the rheology of concentrated soft sphere suspensions in relation to their equilibrium phase diagram.

## 2. The Model

*2.1. The Shear Thickening in Hard Sphere Suspensions.* Hard spheres have an interaction potential that is zero when particles do not overlap and infinite otherwise (Born-repulsion). The phase diagram depends on the volume fraction $\Phi$ of the bare particles determined by

$$\Phi = \frac{\pi a^3 N}{6V}, \tag{1}$$

where $a$ is the particle radius and $N$ the number of particles in the volume $V$. As shown by simulations, monodisperse hard sphere suspensions form a liquid phase and a (face-centred cubic) solid phase for volume fractions $\Phi > 0.54$ in the absence of flow. However, even for a polydispersity of the particles >5%, this transition is suppressed. At a packing fraction $\Phi > 0.58$, the relaxation time becomes large compared to typical experimental time scales. The system does not relax anymore. This jammed state is called a colloidal glass [25, 26]. Approaching the jamming volume fraction $\Phi_j$, the apparent shear viscosity of a hard sphere suspension $\eta_{\mathrm{HS}}$ diverges with

$$\eta_{\mathrm{HS}}(\Phi) \sim \frac{1}{\left(\Phi - \Phi_j\right)^\beta} \tag{2}$$

while $\beta \approx 2$ is the critical exponent for low shear rates $\dot{\gamma} \to 0$ [27]. The mode-coupling theory predicts that a dynamic glass transition occurs already at $\Phi \approx 0.516$. It suggests that approaching the jamming volume fraction large density fluctuations with glass-like dynamics occur in concentrated suspensions. They determine the internal relaxation time of a concentrated suspension and are related to the divergence of the viscosity approaching the jamming transition by

$$\tau_{\mathrm{HS}} \sim \eta_{\mathrm{HS}}(\Phi) \sim \left(\Phi - \Phi_j\right)^{-\beta}. \tag{3}$$

The flow properties of sheared dense suspensions are essentially governed by $\tau_{\mathrm{HS}}$. If the applied shear rate is much smaller than the inverse relaxation time $\dot{\gamma}\tau_{\mathrm{HS}} \ll 1$, density fluctuations disappear before they can be perturbed. For the case $\dot{\gamma}\tau_{\mathrm{HS}} \approx 1$, however, they are deformed by the convective shear flow. As a consequence, density fluctuations are compressed along the compression axis and stretched along the elongation axis of the sheared suspension while rotating in time as schematically displayed in Figure 1. This effect causes a decrease of the apparent viscosity of hard sphere suspensions termed shear thinning [28].

The deformation of density fluctuations leads to the compression of particles arranged along the compression axis. To move away from each other, they must overcome the viscous drag forces created by the small gaps between neighbouring particles. This lubrication effect determines the characteristic separation time $\tau_P$ of a pair of particles. For shear rates $\dot{\gamma}\tau_p \ll 1$, there is sufficient time for hard spheres to detach. However, for $\dot{\gamma}\tau_p \gg 1$, particles arranged along the compression axis start to form transient clusters (so-called hydroclusters). Shear thickening occurs in this view when the applied shear rate is $\dot{\gamma}_C^{\text{HS}}\tau_p \approx 1$ [8]. Based on this approach, several relations have been established characterizing the critical shear stress [2]. As a rule of thumb, the onset of shear thinking can be estimated for Brownian colloids by the relation between Brownian and hydrodynamic forces governed by a Peclet number $\text{Pe} = 6\pi\eta_S a^3 \dot{\gamma}/k_B T$, where $k_B$ is the Boltzmann constant, $T$ the temperature, and $\eta_S$ the Newtonian viscosity of the solvent. Shear thickening occurs for hard spheres when the shear stress $\sigma$ increases thermal diffusion of the particles when $\text{Pe} > 100$. Hence, the critical stress for the onset of shear thickening is of the order [2]

$$\sigma_C^{\text{HS}} = \dot{\gamma}_C^{\text{HS}}\eta_S \approx \frac{50 k_B T}{3\pi a^3}. \tag{4}$$

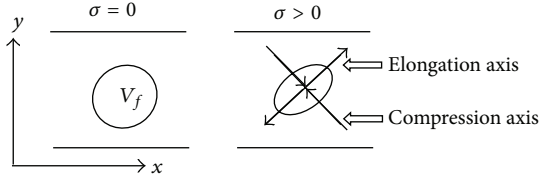

Figure 1: Displayed is the deformation of a large density fluctuation of volume $V_f$ in a simple shear geometry for an applied shear stress $\sigma$.

Shear thickening in hard sphere suspensions occurs as a gradual increase of the viscosity with increasing shear rate (continuous shear thickening), since lubrication forces prohibit a direct contact of the particles. Discontinuous shear thickening, related to a rapid increase of the viscosity at the critical shear rate $\dot{\gamma}_C^{\text{HS}}$, occurs only when the particles come sufficiently close such that lubrication breaks down and the surface roughness of the particles comes into play [18].

### 2.2. The Shear Thickening in Soft Sphere Suspensions

*2.2.1. The Activation Model.* The hard sphere model is an idealization of the colloidal particle interaction. In practical applications mainly soft sphere suspensions with electrostatic or steric repulsion are utilized. For convenience, we want to confine our considerations to soft sphere suspensions of electrically stabilized monodisperse colloidal particles. The DLVO theory states that the total two-particle interaction potential can be expressed as the sum of the double-layer potential $U_{\text{el}}$ and van der Waals attraction $U_{\text{vdW}}$ (note that for high volume fractions short range surface forces and also the surface roughness may come into play not taken into account here; for more details, see [20]):

$$U(h) = U_{\text{el}}(h) + U_{\text{vdW}}(h), \tag{5}$$

where $h$ is the surface-to-surface distance. In numerous cases, a simple equation derived by Hogg et al. [29] was found to be a suitable approximation, which for our case can be expressed for a constant surface potential by

$$U_{\text{CP}}(h) = 2\pi\varepsilon_0\varepsilon_r a \zeta^2 \ln\left(1 + e^{-\kappa h}\right) \tag{6}$$

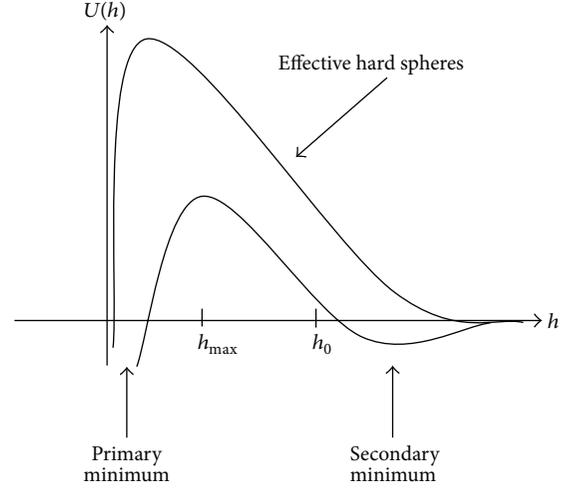

Figure 2: Displayed are two interaction potentials $U(h)$ of electrically stabilized colloidal particles as a function of the surface-to-surface distance $h$. The combination of electrostatic repulsion and van der Waals attraction leads to the occurrence of a primary and a secondary potential minimum, while $h_{\max}$ indicates the potential maximum and $h_0$ the average mutual distance of the particles. For highly stabilized particles (effective hard spheres), the secondary minimum disappears.

and for a constant surface charge approach of the particles by

$$U_{\text{CC}}(h) = -2\pi\varepsilon_0\varepsilon_r a \zeta^2 \ln\left(1 - e^{-\kappa h}\right), \tag{7}$$

where we approximated the surface potential by the $\zeta$-potential. The parameters $\varepsilon_0$ and $\varepsilon_r$ are the absolute and relative dielectric constants. The Debye reciprocal length $\kappa$ is defined by

$$\kappa = \sqrt{\frac{2 C_S N_A Z^2 e_0^2}{\varepsilon_0 \varepsilon_r k_B T}} \tag{8}$$

while $e_0$ is the elementary electric charge, $N_A$ the Avogadro number, $Z$ the ionic charge number, and $C_S$ the salt concentration. The nonretarded van der Waals attraction between two spheres can be described by

$$U_{\text{vdW}}(h) = -\frac{Aa}{12h}, \tag{9}$$

where the effective Hamaker constant $A$ is determined by the dielectric constants of the solvent-particle combination. The corresponding interaction potential is schematically displayed in Figure 2. It consists of a primary and a secondary minimum due to van der Waals attraction and a repulsive potential barrier caused by the electric double layer.

While the hard sphere phase diagram is a function of the volume fraction, the phase diagram of soft spheres is much richer. The two-particle interaction potential depends on many variables characterizing the impact of attractive and repulsive forces. We want to confine the discussion here to two variables: the volume fraction $\Phi$ and the Debye screening reciprocal length $\kappa$. The latter characterizes the electrostatic interaction while an increasing $\kappa$ indicates decreasing electrostatic repulsion. The equilibrium phase diagram is schematically displayed in Figure 3 [30–32]. The phase diagram of electrically stabilized monodisperse colloidal particles separates into colloidal vapour, liquid, and solid phase (the glass state arises for polydisperse particles). For high $\kappa$ and low $\Phi$, Brownian particles with an attractive interaction form a vapour phase that undergoes a flocculation (gel) transition into a solid with increasing volume fraction. The binodal lines of this first order phase transition are displayed in Figure 3. With decreasing $\kappa$, the vapour phase becomes more stable and turns into a liquid phase. In this phase, Brownian particles have to overcome an increasing electrostatic potential barrier. The transition into (fcc-) colloidal crystal occurs at volume fraction $\Phi_C$. This first order liquid-solid transition is termed coagulation. At high volume fractions, however, suspensions do not exhibit a fast phase separation due to the slow dynamics of the bare (hard sphere) particles. Instead, the particles are captured in metastable glassy states.

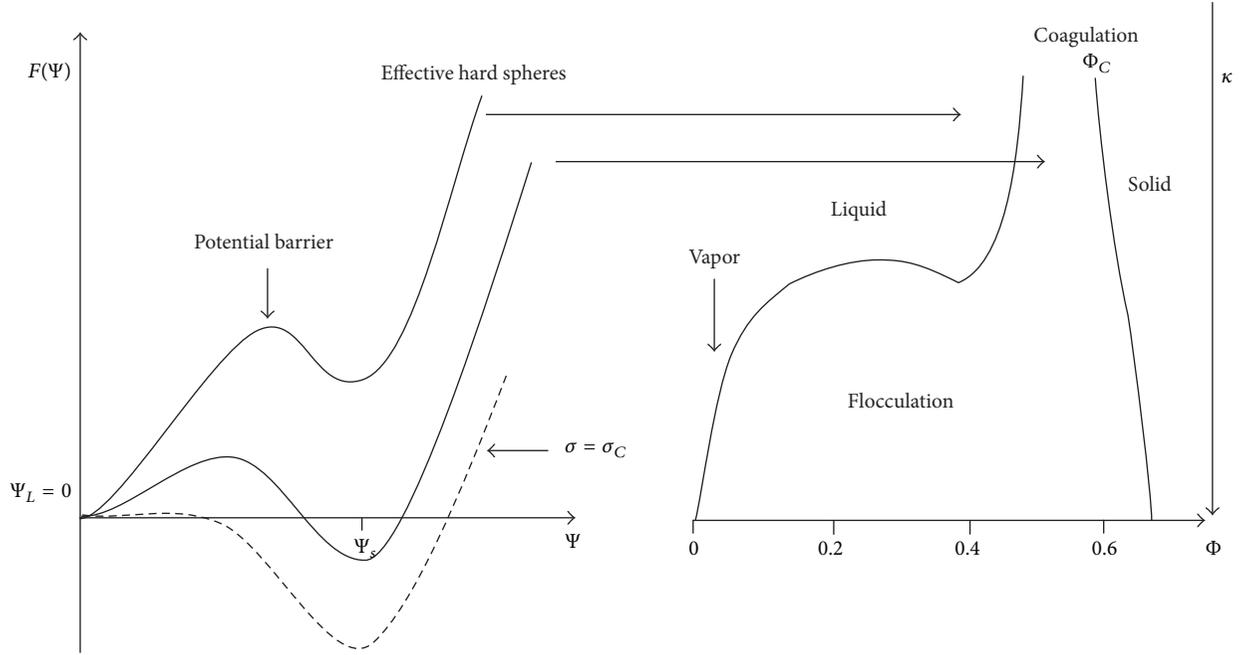

Figure 3: Schematically displayed are the Landau free energies of a stabilized dense suspension in equilibrium near the liquid-solid (coagulation) transition (solid line) and at the critical shear stress $\sigma = \sigma_C$ (dashed line). Also displayed schematically is the equilibrium phase diagram of electrically stabilized colloidal suspensions as a function of the volume fraction $\Phi$ and Debye screening parameter $\kappa$ [30–32]. It shows the first order transition lines separating the colloidal vapor, liquid, and solid phase.

The coagulation transition induced by Brownian motion at low volume fractions (also termed perikinetic coagulation) can be described as an activation process determined by the frequency $\nu$ of bond-forming events. In equilibrium, particles may overcome their repulsion with the frequency $\nu$ by thermal excitations. The nucleation rate is governed for a pair of colloidal particles by

$$\nu = \nu_0 \exp\left(-\frac{U_B}{k_B T}\right), \qquad (10)$$

where $U_B$ is the repulsive energy barrier and $\nu_0$ the collision frequency. The potential barrier formed by the two-particle interaction has the form

$$U_B = U(h_{\max}) - U(h_0) \qquad (11)$$

while $h_0$ is the equilibrium particle surface-to-surface distance and $h_{\max}$ expresses the distance to the maximum of the interaction potential (Figure 2).

In sheared suspensions, colloidal particles arranged along the compression axis are pushed together. The key idea to understand shear thickening in soft sphere suspensions is to realize that these particles may also overcome their mutual repulsion and form transient bonds. For a sheared suspension, we have to correct the undisturbed activation model equation (10) by taking a perturbation due to the applied shear stress $\sigma$ into account. For a pair of colloidal particles arranged along the compression axis, the perturbed frequency can be written as [19]

$$\nu \sim \exp\left(-\frac{U_B - \sigma V^*}{k_B T}\right), \qquad (12)$$

where the activation volume $V^*$ is of the order of the free volume per particle:

$$V^* = \frac{4}{3}\pi a^3 \frac{\Phi_C}{\Phi}. \qquad (13)$$

The critical shear stress at which the frequency becomes a maximum is determined by

$$\sigma_C = \frac{U_B}{V^*} = \eta \dot{\gamma}_C. \qquad (14)$$

The critical stress is a function of the repulsive potential barrier and the free volume. It scales with $a^{-3}$ like (4) for hard spheres. The nucleation rate can be rewritten with (14) as

$$\nu \sim \exp\left(-\frac{\sigma_C - \sigma}{k_B T}\right). \tag{15}$$

The critical shear stress $\sigma_C$ can be also expressed as the product of a viscosity $\eta$ and a critical shear rate $\dot{\gamma}_C$. Since the critical shear stress is nearly independent of the volume fraction (see below), the corresponding critical shear rate is governed by the divergence of the viscosity due to lubrication forces with increasing volume fraction $\eta(\Phi) \sim \eta_{HS}(\Phi)$. The critical shear rate for soft spheres is therefore determined by

$$\dot{\gamma}_C(\Phi) = \frac{\sigma_C(\Phi)}{\eta(\Phi)} \approx C_O (\Phi - \Phi_C)^\beta, \tag{16}$$

where $C_O$ is a free parameter.

For volume fractions $\Phi > \Phi_p$ (where $\Phi_p$ is the percolation volume fraction), clusters of aggregated colloidal particles may even span the entire suspension and form jammed force chains along the compression axis of the sheared suspension. These force chains can cause dilation of the sheared suspension at $\sigma_C$, often observed in connection with discontinuous shear thickening [2].

*2.2.2. Shear Thickening as a Shear Induced Phase Transition.* Differently from the two-particle approach of the activation model, in this section a continuous model is established. For this purpose, we first determine the rheological properties of dense suspensions in a hydrodynamic approach. A colloidal suspension close to a phase transition can be described by a Landau model and is termed here as near-critical suspension. Taking advantage from these models, the critical shear stress for the occurrence of aggregated colloidal particles is obtained from a subcritical bubble approach.

*(1) The Hydrodynamic Model.* We want to specify the rheological properties of a near-critical suspension by applying a two-fluid model. The suspending medium of volume fraction $1 - \Phi$, with density $\rho_s$ and viscosity $\eta_S$, is treated as the first fluid. The second fluid is a continuous medium formed by the ensemble of colloidal particles of volume fraction $\Phi$, density $\rho_p$, and viscosity $\eta_P$. The mass densities of the two fluids, $\rho_s$ and $\rho_p$, are convected by the velocities $\mathbf{v}_p$ and $\mathbf{v}_S$ governed by the conservation relations:

$$\frac{\partial \rho_S}{\partial t} = -\nabla(\rho_S \mathbf{v}_S); \quad \frac{\partial \rho_P}{\partial t} = -\nabla(\rho_P \mathbf{v}_P). \tag{17}$$

The average velocity $\mathbf{v}$ is defined by

$$\mathbf{v} = \frac{1}{\rho}(\rho_S \mathbf{v}_S + \rho_P \mathbf{v}_P), \tag{18}$$

where total density reads

$$\rho = \rho_S + \rho_P. \tag{19}$$

The equations of motion for the two fluids are

$$(1 - \Phi)\rho_S \frac{\partial \mathbf{v}_S(\mathbf{r}, t)}{\partial t} = -\nabla p_S(\mathbf{r}, t) + \nabla \boldsymbol{\sigma}_S(\mathbf{r}, t) - \Xi \mathbf{w},$$
$$\Phi \rho_P \frac{\partial \mathbf{v}_P(\mathbf{r}, t)}{\partial t} = -\nabla p_P(\mathbf{r}, t) + \nabla \boldsymbol{\sigma}_P(\mathbf{r}, t) + \Xi \mathbf{w}, \tag{20}$$

where $p_P$, $p_S$ and $\boldsymbol{\sigma}_p$, $\boldsymbol{\sigma}_S$ are the pressure and viscous stress tensor of the colloidal particle and solvent medium, respectively. The velocity difference between the two fluids is

$$\mathbf{w} = \mathbf{v}_S - \mathbf{v}_p. \tag{21}$$

The friction coefficient can be approximated for near-critical suspensions by

$$\Xi \sim 6\pi \eta_S \xi, \tag{22}$$

where $\xi$ is the correlation length of density fluctuations formed by colloidal particles. Since $\xi$ is a large quantity for near-critical suspensions, we confine the discussion here to the limit of strong coupling $\mathbf{w} \approx 0$. In this case, the continuity equation of the suspension has the usual form

$$\frac{\partial \rho(\mathbf{r}, t)}{\partial t} = -\nabla(\rho(\mathbf{r}, t) \mathbf{v}(\mathbf{r}, t)), \tag{23}$$

where the mean velocity is given by (18). However, the viscosity of dense suspensions diverges with increasing volume fraction as suggested by (2). It causes a viscoelastic response of dense suspensions. This effect can be taken into account by applying a Maxwell model for the momentum relation with the constitutive equation [33]

$$\frac{\partial \sigma(\mathbf{r}, t)}{\partial t} + \frac{1}{\tau}\sigma(\mathbf{r}, t) = G\frac{\partial \gamma(\mathbf{r}, t)}{\partial t} \tag{24}$$

while $\gamma$ is a shear deformation and $G$ an effective shear modulus. The mechanical properties of a viscoelastic medium depend on the relaxation time $\tau$ given for a dense suspension by (3). For $\Phi \rightarrow 0$, the suspension behaves as a viscous fluid because $\tau \rightarrow 0$. For large volume fractions, however, the relaxation time diverges ($\tau \rightarrow \infty$) and the suspension responds as an elastic solid.

*(2) The Landau Model.* The phase diagram displayed in Figure 3 suggests that a first order liquid-solid coagulation transition occurs in concentrated monodisperse soft sphere suspensions at rest. In order to describe this phase transition, we take advantage from a Landau model [22]. For this purpose, an order parameter characterizing the density difference between the average density of the colloidal liquid phase $\rho_0$ and the local density can be defined by

$$\Psi(\mathbf{r}, t) = \rho(\mathbf{r}, t) - \rho_0, \tag{25}$$

where the vector $\mathbf{r}$ indicates the spatial location. The spatially averaged order parameter $\langle \Psi \rangle$ has the property to be zero in the liquid phase and nonzero in the coagulated solid phase. In thermal equilibrium, the free energy density can be established as a Taylor expansion of the order parameter around the instability. Expanding the free energy up to the forth order in $\Psi(\mathbf{r}, t)$, we obtain the standard Ginzburg-Landau free energy scaled by the thermal energy $k_B T$ [23]:

$$F(\Psi) = \frac{1}{k_B T} \int \left( \frac{\iota}{2} |\nabla \Psi(\mathbf{r}, t)|^2 + \frac{\alpha}{2} \Psi(\mathbf{r}, t)^2 + \frac{\lambda}{3} \Psi(\mathbf{r}, t)^3 + \frac{\chi}{4} \Psi(\mathbf{r}, t)^4 \right) d^3 r. \quad (26)$$

The first term takes contributions due to spatial variations of the order parameter into account and $\iota$, $\alpha$, $\lambda$, and $\chi$ are free parameters. For simplicity, the first order coagulation transition is treated as weakly first order such that $\iota$, $\chi > 0$, $\lambda < 0$, and $|\lambda| \ll |\alpha|, |\chi|$. It implies that we confine our considerations to near-critical colloidal liquid with small $\kappa$ (Figure 3).

The free energy minimum of the spatially averaged order parameter $\langle \Psi \rangle$ corresponds to

$$\langle \Psi \rangle \cong \begin{cases} 0 & \text{for } \alpha \geq \dfrac{\lambda^2}{2\chi} \\ -\dfrac{\lambda}{2\chi} \pm \sqrt{\left(\dfrac{\lambda}{2\chi}\right)^2 - \dfrac{\alpha}{\chi}} & \text{for } \alpha < \dfrac{\lambda^2}{2\chi}. \end{cases} \quad (27)$$

For $\alpha \gg 0$, the liquid state is stable, since the free energy has a minimum at $\langle \Psi \rangle = \Psi_L = 0$. For $\alpha \ll 0$, the order parameter becomes nonzero and describes the solid phase with a higher stationary mean density $\langle \Psi \rangle$, while $\Psi_S^2 \approx \alpha/\chi$. The coagulation transition into a solid occurs when $\alpha(\Phi, \kappa) \approx 0$. Hence, the parameter $\alpha$ can be expanded near the liquid-solid transition as

$$\alpha(\Phi, \kappa) \cong \alpha' (\Phi - \Phi_C)(\kappa_C - \kappa), \quad (28)$$

where $\kappa_C$ indicates a critical Debye screening parameter and $\alpha'$ is a free parameter.

Near-critical suspensions are characterized by large order parameter fluctuations with a characteristic size that can be estimated by the correlation length $\xi \approx (\partial^2 F / \partial \langle \Psi \rangle^2)^{-1/2}$. (For a second order transition, the correlation length diverges and the fluctuations become scale invariant, such that no characteristic length exists. However, for first order transitions, the correlation length remains finite.) The time evolution of the order parameter fluctuations is determined by [23]

$$\frac{d\Psi(\mathbf{r}, t)}{dt} = -\Gamma \nabla^2 \frac{\delta F(\Psi)}{\delta \Psi(\mathbf{r}, t)}, \quad (29)$$

where $\Gamma$ is a kinetic coefficient, $\delta$ indicates a variational derivation of the free energy, and we used that the order parameter $\Psi$ is conserved. For the lowest order contribution of the order parameter, we obtain in the Fourier-space with wave vector $\mathbf{k}$

$$\frac{\partial \Psi(\mathbf{k}, t)}{\partial t} \cong -\Gamma k^2 \left( \alpha + k^2 \iota \right) \Psi(\mathbf{k}, t). \quad (30)$$

The relaxation time of order parameter fluctuations becomes, therefore,

$$\tau \sim \frac{1}{\Gamma k^2 \alpha(\Phi, \kappa)}. \quad (31)$$

Obviously the relaxation time of the density fluctuations becomes large for $\alpha(\Phi, \kappa) \to 0$. This effect is known as critical slowing-down.

We have to keep in mind that in a dense suspension internal deformations relax slowly. This effect can be taken into account by including a coupling between the order parameter $\Psi(\mathbf{r}, t)$ and internal deformations of the strain component $\gamma(\mathbf{r}, t)$. The free energy equation (26) becomes

$$F(\Psi, \gamma) \cong \frac{1}{k_B T} \int \left( \frac{\iota}{2} |\nabla \Psi(\mathbf{r}, t)|^2 + \frac{\alpha}{2} \Psi(\mathbf{r}, t)^2 \right.$$
$$+ \frac{\lambda}{3} \Psi(\mathbf{r}, t)^3 + \frac{\chi}{4} \Psi(\mathbf{r}, t)^4$$
$$\left. + \frac{G}{2} \gamma(\mathbf{r}, t)^2 - \theta \gamma(\mathbf{r}, t) \Psi(\mathbf{r}, t)^2 \right) d^3 r, \quad (32)$$

where the fifths term is the contribution of a shear perturbation to the free energy. The last term expresses the lowest order coupling between a shear deformation component $\gamma(\mathbf{r}, t)$ and the order parameter $\Psi(\mathbf{r}, t)$, with the coupling constant $\theta > 0$. The lowest order coupling must be proportional to $\Psi^2$ [34] (a shear deformation cannot generate a nonzero mean order parameter). In equilibrium, the magnitude of the averaged strain component $\langle \gamma \rangle$ can be obtained from

$$\frac{\partial F(\Psi, \langle \gamma \rangle)}{\partial \langle \gamma \rangle} = G \langle \gamma \rangle + \theta \langle \Psi \rangle^2 = 0. \quad (33)$$

In the liquid phase, the mean order parameter is $\langle \Psi \rangle = 0$. As expected, the mean deformation disappears in a colloidal liquid at rest $\langle \gamma \rangle = 0$, since $G \neq 0$.

*(3) The Subcritical Bubble Model.* We want to study a near-critical dense soft-sphere suspension in a simple shear geometry perturbed by a uniform shear flow in the $x$-direction: $v_x = y\dot{\gamma}$, where internal shear deformations have the form $\gamma_{xy} = \partial u_x / \partial y$ (Figure 1). When a continuous shear flow is imposed to the suspension, the order parameter kinetics can be described by a convection-diffusion equation of the form [23]

$$\frac{d\Psi(\mathbf{r}, t)}{dt} + \nabla \left( \Psi(\mathbf{r}, t) v_x(\mathbf{r}, t) \right) = -\Gamma \nabla^2 \frac{\delta F(\Psi)}{\delta \Psi(\mathbf{r}, t)}. \quad (34)$$

It suggests that density fluctuations are convected during their lifetime by the applied shear stress $\sigma_{xy}$ as displayed in Figure 1. However, as suggested by the hydrodynamic model, a sheared dense suspension generates stationary internal shear deformations:

$$\gamma_{xy}(\mathbf{r}, t) \cong G^{-1} \sigma_{xy}(\mathbf{r}, t). \quad (35)$$

That means that a considerable contribution of the applied shear stress is not immediately dissipated but induces internal deformations.

The key idea of this consideration is to estimate the chance for the occurrence of fluctuations of the solid (coagulated) phase within a sheared suspension in the liquid phase. For this purpose, we take advantage from a subcritical bubble approach [34]. The subcritical bubble model is based on the idea that subcritical bubbles can be approximately described in equilibrium as symmetric Gaussian-shaped spheres with radius $R$. The bubble can be parameterized as [35]

$$\Psi(r) = \Psi_{\text{Core}} \exp\left(-\frac{r^2}{R^2}\right) \quad (36)$$

while the core value of the order parameter is $\Psi_{\text{Core}} \approx \Psi_S$. The nucleation rate of subcritical bubbles of coagulated particles of volume $V(R)$ generated in a sheared suspension can be estimated from

$$\nu \sim \exp\left(-\frac{1}{k_B T} \int_{V(R)} F(r, \Psi_{\text{Core}}, \gamma_{xy}) d^3 r\right). \quad (37)$$

Since the initial shape of a subcritical bubble can be approximately given by (36), we obtain for the lowest order in $\Psi_{\text{Core}}$ a nucleation rate:

$$\nu \sim \exp\left(-\frac{(\alpha - \theta \langle \gamma_{xy} \rangle) \Psi_{\text{Core}}^2}{k_B T}\right). \quad (38)$$

Because the mean strain-field in a sheared dense suspension is not zero but of the order

$$\langle \gamma_{xy} \rangle \cong \frac{\sigma_{xy}}{G}, \quad (39)$$

the nucleation rate has a maximum at a critical shear stress:

$$\sigma'_C = \frac{G\alpha}{\theta}. \quad (40)$$

Therefore, the nucleation rate of subcritical bubbles consisting of a few coagulated particles can be written as

$$\nu \sim \exp\left(-\frac{\sigma'_C - \sigma_{xy}}{k_B T}\right). \quad (41)$$

The generation rate of subcritical bubbles is formally equivalent to the nucleation rate equation (15). It suggests that the critical shear stress related to the occurrence of small subcritical bubbles can be approximated by the two-particle critical stress established by the activation model $\sigma'_C \approx \sigma_C$. Since subcritical bubbles (coagulated clusters) increase the viscosity, this transition is accompanied by shear thickening. For decreasing volume fractions $\Phi \to 0$, the relaxation time vanishes: $\tau \to 0$. As a consequence the shear thickening effect disappears in low concentrated suspensions because internal deformations are very small $\langle \gamma \rangle \sim \dot{\gamma}\tau \approx 0$. In this case, the nucleation rate of coagulated particle clusters equation (38) disappears.

The model gives also an explanation for the occurrence of reversible and irreversible shear thickening (orthokinetic coagulation [36]). As displayed in Figure 3, the Landau free energy of a suspension exhibiting irreversible shear thickening must have a stable minimum at $\Psi_S$ (at $\sigma = 0$). Applying a shear stress, the potential barrier of the free energy effectively decreases such that at $\sigma_C$ the nucleation rate is sufficiently high to form subcritical bubbles of coagulated particles. Ceasing the shear stress, the coagulated structure cannot relax into the liquid state when the potential barrier increases the thermal energy. In the case of reversible shear thickening, however, it is known that for any applied shear stress no coagulated structure is evident after ceasing the stress [37]. The free energy must have therefore an unstable minimum at $\Psi_S$ (at $\sigma = 0$) with a potential barrier that is much less than the thermal energy. Therefore, thermal excitations are sufficient to break up the bonded structure. Colloidal particles have in this case an effective hard sphere interaction potential (Figure 2).

Though van der Waals attraction diverges for $h \to 0$ such that particles should be always bonded in the primary minimum ones, they are captured. The explanation for the thermal breakup of the bonds is given in [20]. The point is that the potential maximum $h_{\max}$ for effective hard spheres is only a few Angstrom from the bare particle surface. Since $h_{\max}$ is of the order of the diameter of the atomic constituents, it can be expected that the surface roughness of the particles prohibits a permanent bond.

Note that, for $\alpha < 0$, the colloidal particles are already bounded in the primary minimum and (40) suggests that $\sigma_C < 0$. That means that shear thickening disappears for a coagulated structure and the rheological properties are those of a coagulated solid. That this is the case has been shown experimentally by Barnes [1]. Modifying the chemistry of a stable suspension by adding a large amount of flocculating agents shear thickening disappears.

Rheological experiments indicate that shear thickening in concentrated suspensions exhibits a viscosity hysteresis between increasing and decreasing shear rates. It means that the critical shear stress related to increasing shear rates is much higher than the critical stress associated with decreasing shear rates. The presented theory can explain this finding. If subcritical bubbles of particles with large attraction (high $\kappa$) are formed, their relaxation time is large. Decreasing the shear rate they remain in the bounded state even at the critical shear stress related to increasing shear rates. The model suggests that the hysteresis effect increases with increasing attraction between the particles, which is known from empirical investigations of discontinuous shear thickening.

## 3. Comparison with Experimental Results

Both the activation approach and the presented shear induced transition model suggest the occurrence of shear thickening at a critical shear stress $\sigma_C$. In order to show the applicability of the theory, we want to consider an experimental investigation performed by Maranzano and Wagner on nonaqueous electrically stabilized suspensions [38]. They studied silica particles in THFFA (tetrahydrofurfuryl alcohol), a system designed to minimize the van der Waals attraction. We want to focus here only on samples denoted by HS600, because for these samples the flow curves have been published for a large number of volume fractions. The concentrated suspensions were found to exhibit discontinuous reversible shear thickening for volume fractions $\Phi \geq 0.55$ in a constant stress sweep. Taking advantage from the experimentally characterized data of the colloidal particles, the critical shear stresses for constant potential (dotted line) and constant charge (solid line) obtained from the activation model are displayed in Figure 4, together with the measured critical shear stresses (squares). The experimental critical stresses fall into the predicted range between the two lines, while they are closer to the constant charge critical stress. (This result can be interpreted as a consequence of the double layer dynamics of the electrostatic repulsion. There is obviously not sufficient time at high critical shear rates to establish a constant potential double layer differently from low critical shear rates at high volume fractions.) Note that (4) suggests a critical shear stress $\sigma_C^{HS} \approx 0.6$ Pa which is an order of magnitude below the experimental data. Also displayed in Figure 4 are the experimental data of the critical shear rate of the sample HS600 at the onset of shear thickening (diamonds). The dashed line indicates a fit of (16) with a volume fraction $\Phi_C = 0.6$, $\beta = 2$, and $C_O = 5 * 10^4$. Note that the activation model is applicable not only to concentrated electrically stabilized but also to sterically stabilized colloidal suspensions [39].

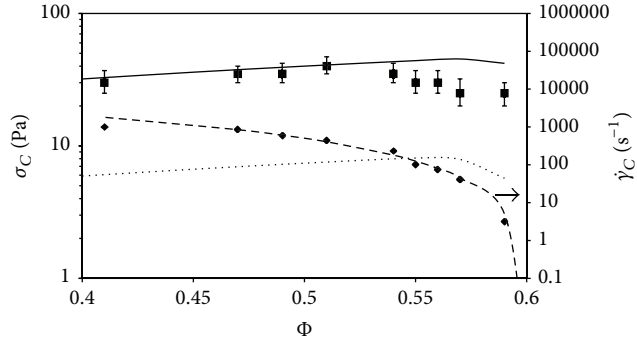

Figure 4: Experimental critical stresses: a function of the volume fraction of the sample HS600 (squares) investigated by Maranzano and Wagner [38]. The solid line indicates the critical stress obtained from the activation model for a constant charge interaction potential and the dotted line with constant potential applying the experimental data $a = 316$ nm, $\zeta = 92.7$, $\kappa a = 87.1$, $k_B = 1.38 * 10^{-23}$ J, $e = 1.6 * 10^{-19}$ As, $\varepsilon_r = 8.2$, $\varepsilon_0 = 8.854 * 10^{-12}$ F/m, $z = 1$, $T = 298$ K, $A = 0.41 * 10^{-21}$ J, $\Phi_C = 0.6$, and $\eta_{THFFA} = 0.005$ Pas. Also displayed are the empirical critical shear rates at the onset of shear thickening (diamonds) of the same sample. The dashed line indicates a fit of (16).

## 4. Conclusion

The paper establishes a model that suggests the occurrence of a shear induced phase transition in dense soft sphere suspensions accompanied with shear thickening. The theory is based on the idea that near-critical suspensions contain large density fluctuations. Because a concentrated suspension can be treated as a viscoelastic medium, the application of a continuous shear flow generates internal deformations. The coupling of these deformations to the order parameter related to the liquid-solid transition induces subcritical bubbles of the solid phase in the stable liquid phase if a continuous shear perturbation is applied. Since subcritical bubbles consist of coagulated particles, their appearance causes an increase of the viscosity at a critical stress (shear thickening).

The flow properties of sheared suspensions are governed by the interaction potential of the particles. This interaction is related to the location in the equilibrium phase diagram. Differently from shear thickening in hard sphere suspensions, the phase diagram of soft spheres depends on many parameters characterizing the mutual interaction. We want to confine the discussion here to electrically stabilized monodisperse suspensions where the volume fraction $\Phi$ and the Debye screening reciprocal length $\kappa$ are treated as free parameters. Displayed schematically in Figure 5 are two paths through the equilibrium phase diagram of electrically stabilized suspensions and the expected rheological response in terms of the viscosity $\eta(\Phi, \kappa)$. The dashed line indicates a path with increasing volume fraction $\Phi$ and constant $\kappa$. In this case, the colloidal particles are stabilized by the electrostatic repulsion and the critical stress is nearly independent of the volume fraction. These suspensions are governed by the divergence of the viscosity with increasing volume fraction caused by lubrication forces. As found experimentally, increasing the volume fraction, the character of the transition turns into discontinuous shear thickening [40].

The second path in Figure 5 is related to suspensions with an increasing Debye reciprocal length $\kappa$ of the colloidal particles and constant $\Phi$. This path is absent in a hard sphere phase diagram. Starting in the liquid phase, the particles are highly stabilized (effective hard spheres) and continuous shear thickening occurs at relatively high shear rates. Increasing $\kappa$ implies that the particle solvent combination approaches the coagulation transition. This can be done, for example, by increasing the salt concentration. As a consequence, the electrostatic repulsion decreases and the critical stress declines. The model suggests that reversible shear thickening turns into irreversible shear thickening with increasing $\kappa$. Finally shear thickening disappears in a stable coagulated structure as known from empirical investigations [41].

For a quantitative comparison, the presented theory is applied to experimental data of shear thickening in dense electrically stabilized suspensions [38]. The good coincidence between the model predictions and the experimentally obtained critical shear stresses confirms this approach.

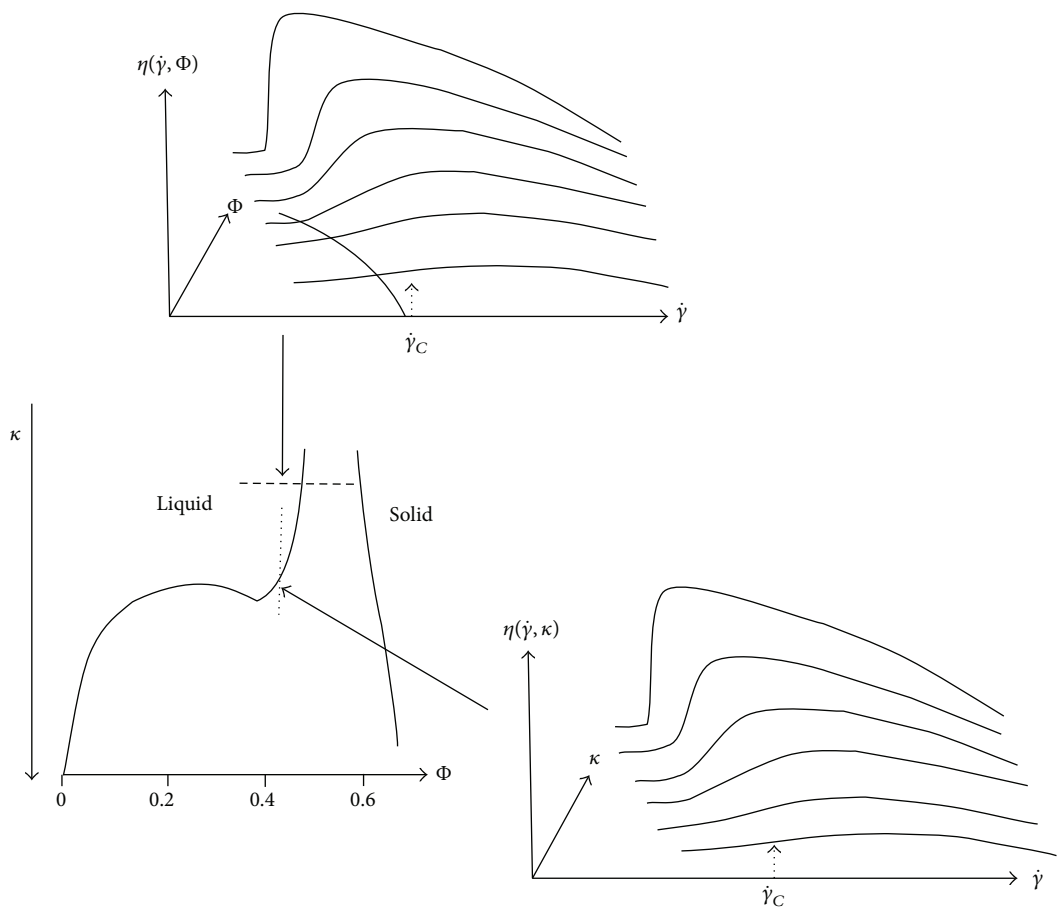

Figure 5: Schematically displayed is the equilibrium phase diagram of electrically stabilized monodisperse colloidal suspensions. The model suggests that shear thickening in soft sphere suspensions depends on the position in the phase diagram. The expected viscosity is displayed for two different paths as a function of the shear rate (logarithmic scaling). The dashed line in the phase diagram indicates a path with varying volume fraction and the dotted line a path with changing the Debye screening parameter $\kappa$. For comparison with empirical data, see, for example, [40, 41].

## Conflict of Interests

The authors declare that there is no conflict of interests regarding the publication of this paper.

## References

[1] H. A. Barnes, "Shear-thickening ('Dilatancy') in suspensions of nonaggregating solid particles dispersed in Newtonian liquids," *Journal of Rheology*, vol. 33, no. 2, pp. 329–366, 1989.

[2] E. Brown and H. M. Jaeger, "Shear thickening in concentrated suspensions: phenomenology, mechanisms and relations to jamming," *Reports on Progress in Physics*, vol. 77, no. 4, Article ID 046602, 2014.

[3] N. J. Wagner and J. F. Brady, "Shear thickening in colloidal dispersions," *Physics Today*, vol. 62, no. 10, pp. 27–32, 2009.

[4] F. Picano, W.-P. Breugem, D. Mitra, and L. Brandt, "Shear thickening in non-Brownian suspensions: an excluded volume effect," *Physical Review Letters*, vol. 111, no. 9, Article ID 098302, 2013.

[5] J. F. Morris, "A review of microstructure in concentrated suspensions and its implications for rheology and bulk flow," *Rheologica Acta*, vol. 48, no. 8, pp. 909–923, 2009.

[6] J. F. Brady and G. Bossis, "The rheology of concentrated suspensions of spheres in simple shear flow by numerical simulation.," *Journal of Fluid Mechanics*, vol. 155, pp. 105–129, 1985.

[7] J. W. Bender and N. J. Wagner, "Reversible shear thickening in monodisperse and bidisperse colloidal dispersions," *Journal of Rheology*, vol. 33, pp. 329–366, 1989.

[8] J. R. Melrose, "Colloid flow during thickening—a particle level understanding for core-shell particles," *Faraday Discussions*, vol. 123, pp. 355–368, 2003.

[9] R. Seto, R. Mari, J. F. Morris, and M. M. Denn, "Discontinuous shear thickening of frictional hard-sphere suspensions," *Physical Review Letters*, vol. 111, no. 21, Article ID 218301, 2013.

[10] R. L. Hoffman, "Discontinuous and dilatant viscosity behavior in concentrated suspensions. II. Theory and experimental tests,"

*Journal of Colloid and Interface Science*, vol. 46, no. 3, pp. 491–506, 1974.

[11] W. H. Boersma, J. Laven, and H. N. Stein, "Shear thickening (dilatancy) in concentrated dispersions," *AIChE Journal*, vol. 36, no. 3, pp. 321–332, 1990.

[12] W. H. Boersma, P. J. M. Baets, J. Laven, and H. N. Stein, "Time-dependent behaviour and wall slip in concentrated shear thickening dispersions," *Journal of Rheology*, vol. 35, no. 6, pp. 1093–1119, 1991.

[13] R. L. Hoffman, "Explanations for the cause of shear thickening in concentrated colloidal suspensions," *Journal of Rheology*, vol. 42, no. 1, p. 111, 1998.

[14] M. E. Cates, J. P. Wittmer, J.-P. Bouchaud, and P. Claudin, "Jamming, force chains, and fragile matter," *Physical Review Letters*, vol. 81, no. 9, pp. 1841–1844, 1998.

[15] E. Bertrand, J. Bibette, and V. Schmitt, "From shear thickening to shear-induced jamming," *Physical Review E*, vol. 66, no. 6, Article ID 060401, 2002.

[16] C. B. Holmes, M. Fuchs, and M. E. Cates, "Jamming transitions in a schematic model of suspension rheology," *Europhysics Letters*, vol. 63, no. 2, pp. 240–246, 2003.

[17] C. B. Holmes, M. E. Cates, M. Fuchs, and P. Sollich, "Glass transitions and shear thickening suspension rheology," *Journal of Rheology*, vol. 49, no. 1, pp. 237–269, 2005.

[18] E. Brown and H. M. Jaeger, "The role of dilation and confining stresses in shear thickening of dense suspensions," *Journal of Rheology*, vol. 56, no. 4, pp. 875–924, 2012.

[19] J. Kaldasch, B. Senge, and J. Laven, "Shear thickening in electrically-stabilized colloidal suspensions," *Rheologica Acta*, vol. 47, no. 3, pp. 319–323, 2008.

[20] J. Kaldasch, B. Senge, and J. Laven, "The impact of non-DLVO forces on the onset of shear thickening of concentrated electrically stabilized suspensions," *Rheologica Acta*, vol. 48, no. 6, pp. 665–672, 2009.

[21] J. Kaldasch, B. Senge, and J. Laven, "Shear thickening in electrically stabilized non-aqueous colloidal suspensions," *Applied Rheology*, vol. 19, no. 2, pp. 23493–23499, 2009.

[22] L. D. Landau and E. M. Lifshitz, *Statistical Physics*, Pergamon Press, Oxford, UK, 2000.

[23] S. Puri and V. Vadhawan, *Kinetics of Phase Transitions*, CRC Press, New York, NY, USA, 2009.

[24] A. Onuki, *Phase Transitions Dynamics*, Cambridge University Press, Cambridge, UK, 2002.

[25] A. J. Liu, S. R. Nagel, W. van Saarloos, and M. Wyart, "The jamming scenario—an introduction and outlook," in *Dynamical Heterogeneities in Glasses, Colloids, and Granular Medi*, L. Berthier, G. Biroli, J.-P. Bouchaud, L. Cipelletti, and W. van Saarloos, Eds., Oxford University Presss, Oxford, UK, 2011.

[26] L. Berthier and G. Biroli, "Theoretical perspective on the glass transition and amorphous materials," *Reviews of Modern Physics*, vol. 83, no. 2, pp. 587–645, 2011.

[27] D. E. Quemada, "Models for rheological behavior of concentrated disperse media under shear," in *Advances in Rheology, Vol. 2: Fluids*, pp. 571–582, Universidad Nacional Autonoma de Mexico, Mexico City, Mexico, 1984.

[28] J. M. Brader, "Nonlinear rheology of colloidal dispersions," *Journal of Physics Condensed Matter*, vol. 22, no. 36, Article ID 363101, 2010.

[29] R. Hogg, T. W. Healy, and D. W. Fuerstenau, "Mutual coagulation of colloidal dispersions," *Transactions of the Faraday Society*, vol. 62, pp. 1638–1651, 1966.

[30] J. Kaldasch, J. Laven, and H. N. Stein, "Equilibrium phase diagram of suspensions of electrically stabilized colloidal particles," *Langmuir*, vol. 12, no. 26, pp. 6197–6201, 1996.

[31] V. Morales, J. A. Anta, and S. Lago, "Integral equation prediction of reversible coagulation in charged colloidal suspensions," *Langmuir*, vol. 19, no. 2, pp. 475–482, 2003.

[32] G. F. Wang and S. K. Lai, "Domains of phase separation in a charged colloidal dispersion driven by electrolytes," *Physical Review E: Statistical, Nonlinear, and Soft Matter Physics*, vol. 70, no. 5, Article ID 051402, 2004.

[33] L. D. Landau and E. M. Lifshitz, *Theory of Elasticity*, Pergamon Press, Oxford, UK, 2000.

[34] J. Kaldasch, B. Senge, and J. Laven, "Structural transitions in sheared electrically stabilized colloidal crystals," *Journal of Applied Chemistry*, vol. 2013, Article ID 909841, 7 pages, 2013.

[35] M. Gleiser, R. C. Howell, and R. O. Ramos, "Nonequilibrium precursor model for the onset of percolation in a two-phase system," *Physical Review E: Statistical, Nonlinear, and Soft Matter Physics*, vol. 65, no. 3, Article ID 036113, 2002.

[36] S. K. Friedlander, *Smoke, Dust, and Haze: Fundamentals of Aerosol Dynamics*, Oxford University Press, New York, NY, USA, 2000.

[37] Y. S. Lee and N. J. Wagner, "Dynamic properties of shear thickening colloidal suspensions," *Rheologica Acta*, vol. 42, no. 3, pp. 199–208, 2003.

[38] B. J. Maranzano and N. J. Wagner, "The effects of particle size on reversible shear thickening of concentrated colloidal dispersions," *The Journal of Chemical Physics*, vol. 114, no. 23, pp. 10514–10527, 2001.

[39] J. Kaldasch and B. Senge, "Shear thickening in polymer stabilized colloidal suspensions," *Colloid and Polymer Science*, vol. 287, no. 12, pp. 1481–1485, 2009.

[40] H. M. Laun, "Rheological properties of polymer dispersions with respect to shear-induced particle structure," in *Progress and Trends in Rheology II*, H. Giesesekus, Ed., pp. 287–290, Steinkopff, Darmstadt, Germany, 1988.

[41] H. M. Laun, "Rheology and particle structures of concentrated polymer dispersions," in *Proceedings of the 10th International Congress on Rheology*, vol. 1, pp. 37–42, 1988.

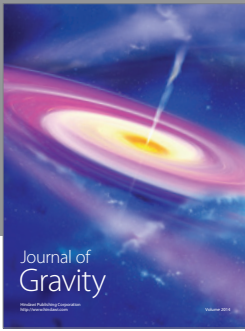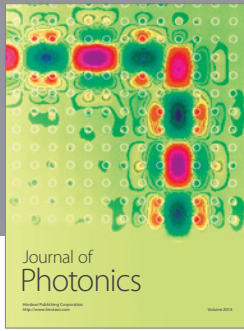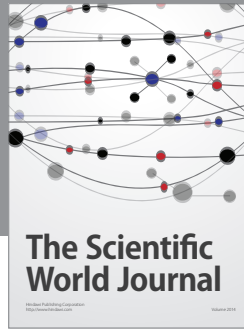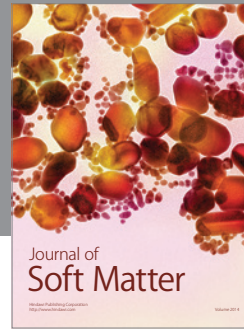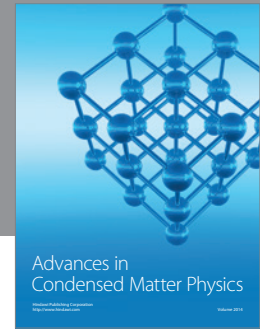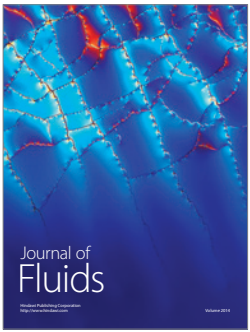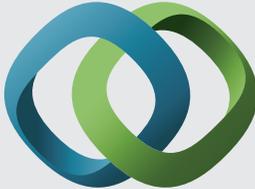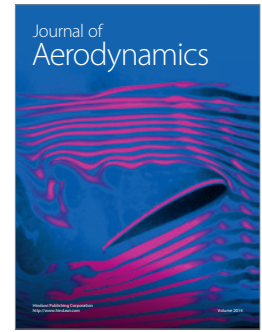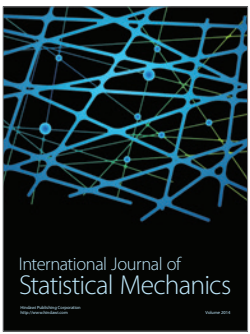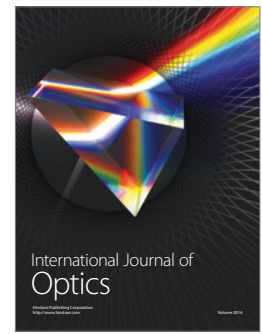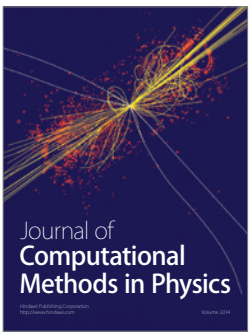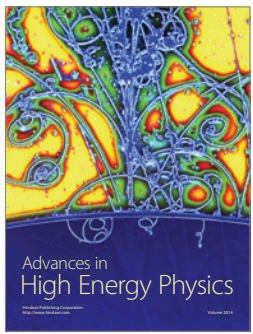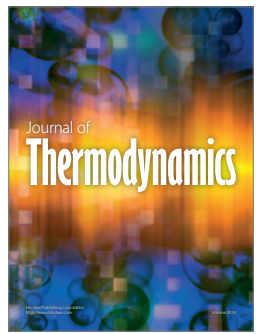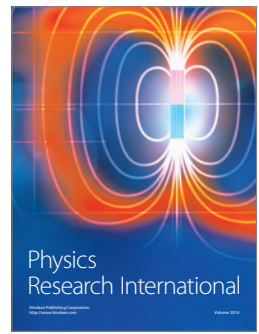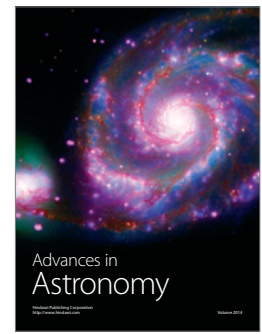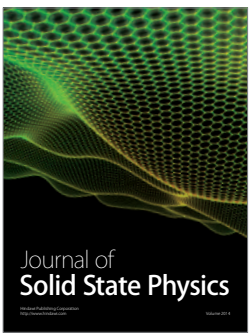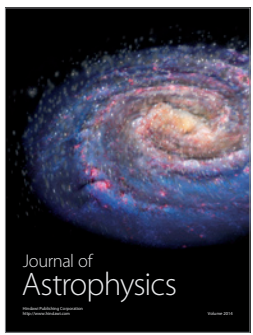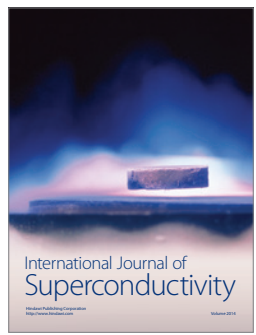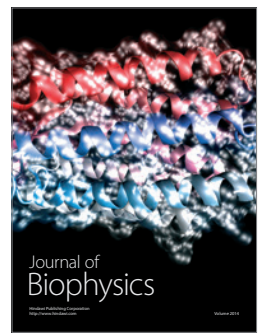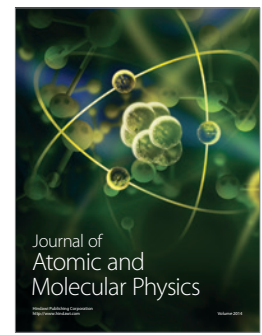